\documentclass[
reprint, twocolumn,
superscriptaddress,
nofootinbib,
amsmath,amssymb, aps, pra,
]{revtex4}

\usepackage{graphicx}% Include figure files
\usepackage{subfigure}
\usepackage{dcolumn}% Align table columns on decimal point
\usepackage{bm}% bold math
\usepackage{color} % change color of text
\usepackage{CJK}
\usepackage{dsfont}
\usepackage[extension=xxx]{hyperref}

\newcommand{\beq}{\begin{equation}}
\newcommand{\eeq}{\end{equation}}
\newcommand{\beqa}{\begin{eqnarray}}
\newcommand{\eeqa}{\end{eqnarray}}

\def\beq{\begin{equation}}

\begin{document}
\title{{\it Ab initio} study of subsurface diffusion of Cu on the H-passivated Si(001) surface}
\author{A. Rodriguez-Prieto}
\affiliation{Departamento de Matematica Aplicada, UPV/EHU}
\author{D. R. Bowler}
\affiliation{Department of Physics and Astronomy, UCL}
%\date{\today}

\begin{abstract}
In this paper we use density-functional theory calculations to analyze both the stability and diffusion of Cu
adatoms near and on the H-passivated Si(001) surface. Two different Cu sources are considered: depositing Cu
from vacuum, and contaminating Cu outdiffusing from bulk Si. Deposited Cu from vacuum quickly moves
subsurface to an interstitial site in the third Si layer (T2). Once there, Cu adatoms enter a subsurface zigzag
migration route between T2 and another subsurface site, T2→HSL→T2, along the dimer row direction.
Contaminating Cu outdiffusing from bulk is found to be a fast diffuser along both parallel and perpendicular
directions to the dimer row when far from the surface. It is attracted to the layers close to the surface and
becomes trapped at an interstitial site located at the sixth Si layer (T3). As the outdiffusing Cu atoms get closer
to the surface, a channeling zigzag diffusion along the dimer row direction, similar to that one followed by
deposited Cu from vacuum, is favoured over diffusion along the perpendicular direction. These results are
consistent with previous experimental work done on similar systems and will motivate further experiments on
the interesting interaction between Cu and Si surfaces.
\end{abstract}

\maketitle
\section{Introduction}
A clear theoretical understanding of the defects which affect
the performance of silicon-based complementary metaloxide
semiconductor devices is still far from complete and
would be of enormous benefit. Copper is a highly detrimental
and very common contaminant in device manufacturing \cite{Foster2007}
especially since its introduction as the metal for interconnects.
It is an extremely fast diffuser in Si, even at room
temperature \cite{Istratov1998} and forms silicides which degrade device characteristics.
When adsorbed on clean Si(001), Cu forms a distinctive
three-lobed linear feature perpendicular to the dimer rows
which orders to form a c(8x8) reconstruction at high
coverages \cite{Liu2000}. On addition of hydrogen to a Cu-dosed surface,
the feature sizes and appearance change, suggesting that the
Cu may be subsurface \cite{Baker2005}, 
though the structure of these features
is still unknown. The adsorption of hydrogen onto Si(001)
follows various stages \cite{Boland1993}: very low coverages lead to single
hydrogen atoms adsorbed on one side of a dimer, giving a
hemihydride \cite{Radny2007}; however, when the coverage goes beyond
more than about 0.1 ML, these single hydrogens rapidly pair
up to form saturated dimers \cite{Boland1991} with one hydrogen for each
dangling bond; adsorption of a full monolayer leads to the
saturated, monohydride surface. This surface, notated
H:Si(001), is generally very well ordered and unreactive; it
can be used as a mask when creating features such as
dangling-bond wires \cite{Bowler2004} or for overgrowth of systems such as
the Bi nanolines which self-assemble on Si(001) \cite{Owen2006}. While the
overgrowth of Bi lines with the noble metals such as Ag and
Au has been considered, Cu has not been used, presumably
because of its tendency to act as a contaminant.
Here, we present a density-functional theory (DFT) study
of the behaviour of Cu at and below H:Si(001), considering
both the low-energy structures formed and the diffusion barriers.
In a related paper we model the interaction of Cu
with Bi nanolines on H:Si(001). Our purpose is to understand
the behaviour of this important metal at the H:Si(001)
surface and possible mechanisms for decoration of the surface
and Bi nanolines. After detailing our methods, we consider
first the stable sites for Cu near H:Si(001), and then
diffusion barriers for Cu starting on the surface and coming
from the bulk, after which we conclude.

\section{Theoretical and Computational Background}
We have used a plane-wave implementation of DFT
(Refs. \cite{Hohenberg1964} and \cite{Kohn1965}) within the local density approximation\cite{Ceperley1980, Perdew1981, Perdew1992}.
All calculations have been performed with the Vienna ab
initio simulation package (VASP) (Refs. \cite{Kresse1993} and \cite{Kresse1996}). Ultrasoft
pseudopotentials (Ref. 18) are used for all the elements considered:
Cu, Si, and H. We use an energy cutoff of 233.781 eV for the
plane waves (which converges energy differences).
To understand the interaction of Cu with the H:Si(001)
surface, we use a slab with two dimer rows, each containing
four dimers (giving an area of 15.36 Å x 15.36 Å) and with
ten layers. This guarantees that the Cu atoms in different
cells will not interact. For related studies involving the interaction
of Cu with Bi nanolines, we must use a computational
cell with a rather different shape: the nanolines require
a cell ten-dimers long and ten-layers deep, which practically
restricts us to a unit-cell one-dimer row wide for most calculations.
We have therefore repeated our calculations in a
cell of this size to check on the effect of cell size and possible
interactions between Cu atoms in neighbouring cells.
Both cells are illustrated in Fig. 1. For both cells, the bottom
two layers are fixed into their bulk positions and terminated
with two H atoms each to simulate bulk conditions, whereas
the surface dimers (pairs of atoms bonded to each other with
one dangling bond) contained within the slab are covered
with a monohydride surface. The square cell contains 16 Si atoms per layer and 48 H atoms in total, giving 208 atoms
while the longer thin cell contains 20 Si atoms per layer and
60 H atoms in total, giving 260 atoms. We use a
Monkhorst-Pack \cite{Monkhorst1976} mesh of 4x4x1 to sample the Brillouin
Zone for the square cell, and 4x2x1 for the long, thin cell;
both of these give energy-difference convergence.
Calculations of barrier heights used the nudged elastic
band method \cite{Henkelman2000} with a single image (which is equivalent to
the dimer method \cite{Henkelman1999} with very close images or the force inversion
method \cite{Tateyama1996} and gives excellent convergence of
energy-barrier heights). For a number of the migration routes
considered, we found that the Si-Si distances were sufficiently
small that the passage of Cu-caused significant rearrangements
of the lattice and extremely large energy barriers.
We label these routes as “n/a” in Sec. IV.

\section{Stable sites for Cu near H:Si(001)}
We consider first the binding energies (BEs) of Cu at different
possible sites below and at the H:Si(001) surface. The
optimized geometries of the most energetically favoured positions
are then studied in detail. The diffusion of Cu on and
below the H-passivated Si(001) surface is described in Sec.
IV, where different migration routes with their corresponding
energy barriers will be examined.
All the candidate sites are illustrated in Fig. 2 with binding
energies given in Table I. We have selected different
adsorption sites both on the surface and below the surface.
On the surface we consider: the cave site C(-4.41 eV), on
the side of a dimer at the trench between dimer rows; the
pedestal site P(-4.05 eV) between two adjacent dimers
along the dimer row; and the site A(-3.96 eV) close to the
Si-H bond of one of the dimers. Below the surface we consider
three tetrahedral sites: T4(-4.49 eV), T3(-4.65 eV),
and T2(-4.60 eV); and four other sites inside the
hexagonal cages: HSL(-4.60 eV), HSL2(-4.62 eV),
HSL3(-4.53 eV), and HSL4(-4.36 eV). An additional adsorption
site below the dimers, BD(-3.81 eV) is also considered.
These energies are summarised in Table I. The T2
and T3 sites show lower binding energies than T4, which
indicates that, despite the well-known property of Cu as a
fast diffuser inside bulk Si, it will tend to be adsorbed close
to the top layers of the H-passivated Si(001) surface; we
explore the attraction mechanism of Cu to defects
elsewhere. Similarly, HSL2 and HSL3 show lower BEs
than HSL4. However, HSL has a BE slightly higher than
HSL2, as does T2 with respect to T3. This indicates that
despite a clear tendency for Cu to be adsorbed close to the
surface, it does not go to the top Si layers. It is also noteworthy
that tetrahedral positions inside bulk Si(T5,T6, ...) show
the same BE as T4 so that the energies of these sites are not
affected by the constrained bottom layers. The most energetically
favoured sites (T3, HSL2, T2, and HSL in descending
energy) are located between the third and the sixth layers.
The next sites in energy are HSL3, T4, C, and HSL4,
whereas the rest of the sites show sufficiently poor energies
to be unimportant and are not considered further. The large
gap in energy which opens between the C and P sites is a key
indicator. The pedestal P site and its close A site already
show energies 0.60 and 0.69 eV higher than T3. Below the
dimers, BD, is even worse, with an energy 0.83 eV higher and, finally, the B site presents a really poor energy due to
the strong Si-Si bond of the dimers.

\begin{center}
%\begin{tabular}{ c c c }
\begin{tabular}{  c  c  c  } 
\hline \hline
 Adsorp. site & $\Delta E$ (eV) & Binding energy (eV) \\ 
\hline
 T3 & 0 & -4.65 \\  
 HSL2& 0.03 & -4.62 \\ 
 T2 & 0.05 & -4.60 \\
 HSL& 0.05 & -4.60 \\ 
 HSL3& 0.12 & -4.53 \\
 T4 & 0.16 & -4.49 \\  
 C& 0.24 & -4.41 \\ 
 HSL4 & 0.29 & -4.36 \\
 P& 0.60 & -4.05 \\ 
 A& 0.69 & -3.96 \\
 BD & 0.83 & -3.82    
\end{tabular}
\end{center}

%%%%%%%%%%%%%%%%%%%%%%%%%%%%%%%%%%%%%%%%%%%%%%%%%%%%%%%%%%%%%%%%
\begin{figure}[]
\centering
{\includegraphics[width=85mm]{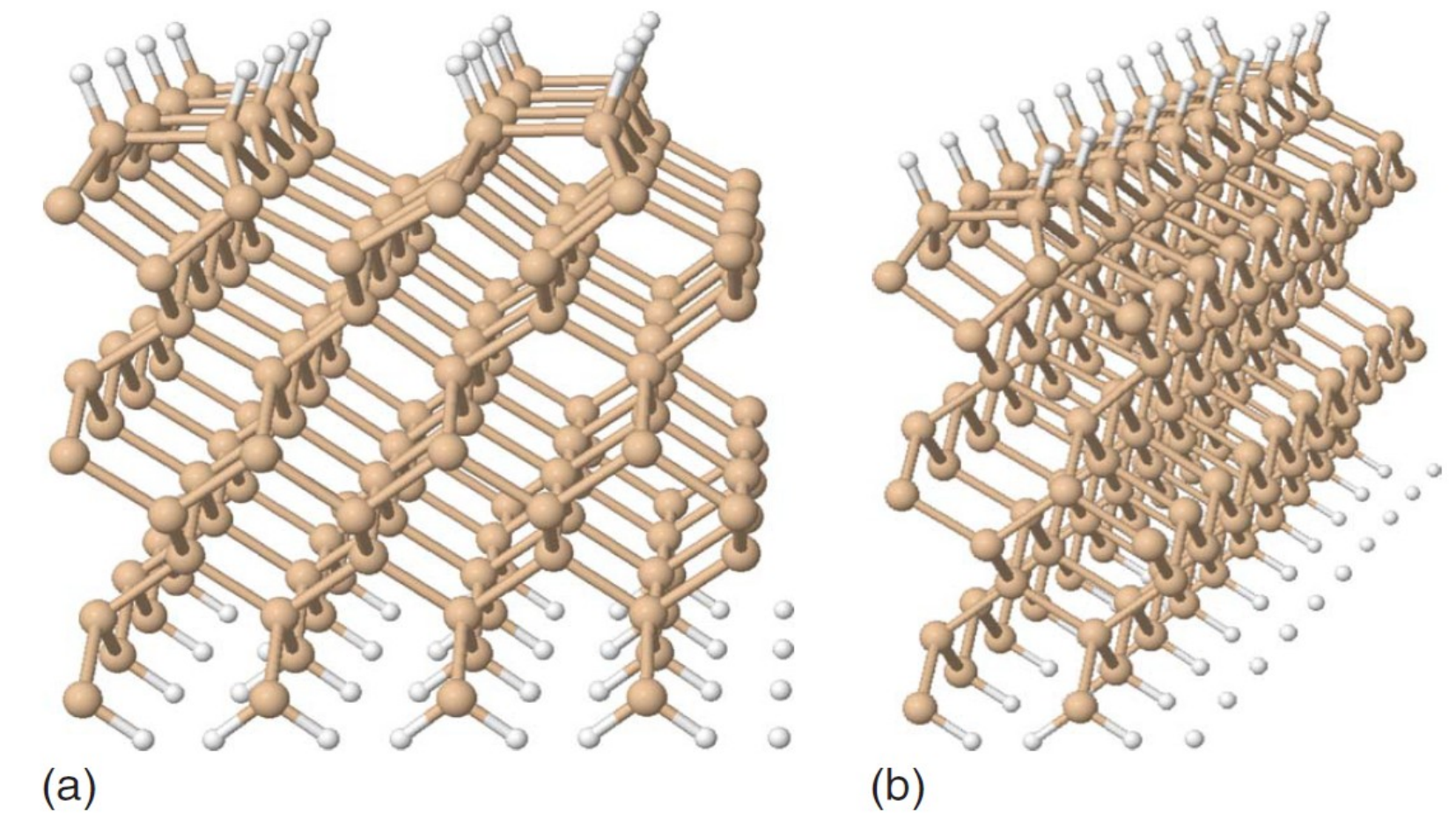}}
\caption{(Color online). Slab geometries used for simulating H-passivated si(001) surface: left, square cell; right, nanoline cell}
\label{fig:1}
\end{figure}
%%%%%%%%%%%%%%%%%%%%%%%%%%%%%%%%%%%%%%%%%%%%%%%%%%%%%%%%%%%%%%

%%%%%%%%%%%%%%%%%%%%%%%%%%%%%%%%%%%%%%%%%%%%%%%%%%%%%%%%%%%%%%%%
\begin{figure}[]
\centering
{\includegraphics[width=85mm]{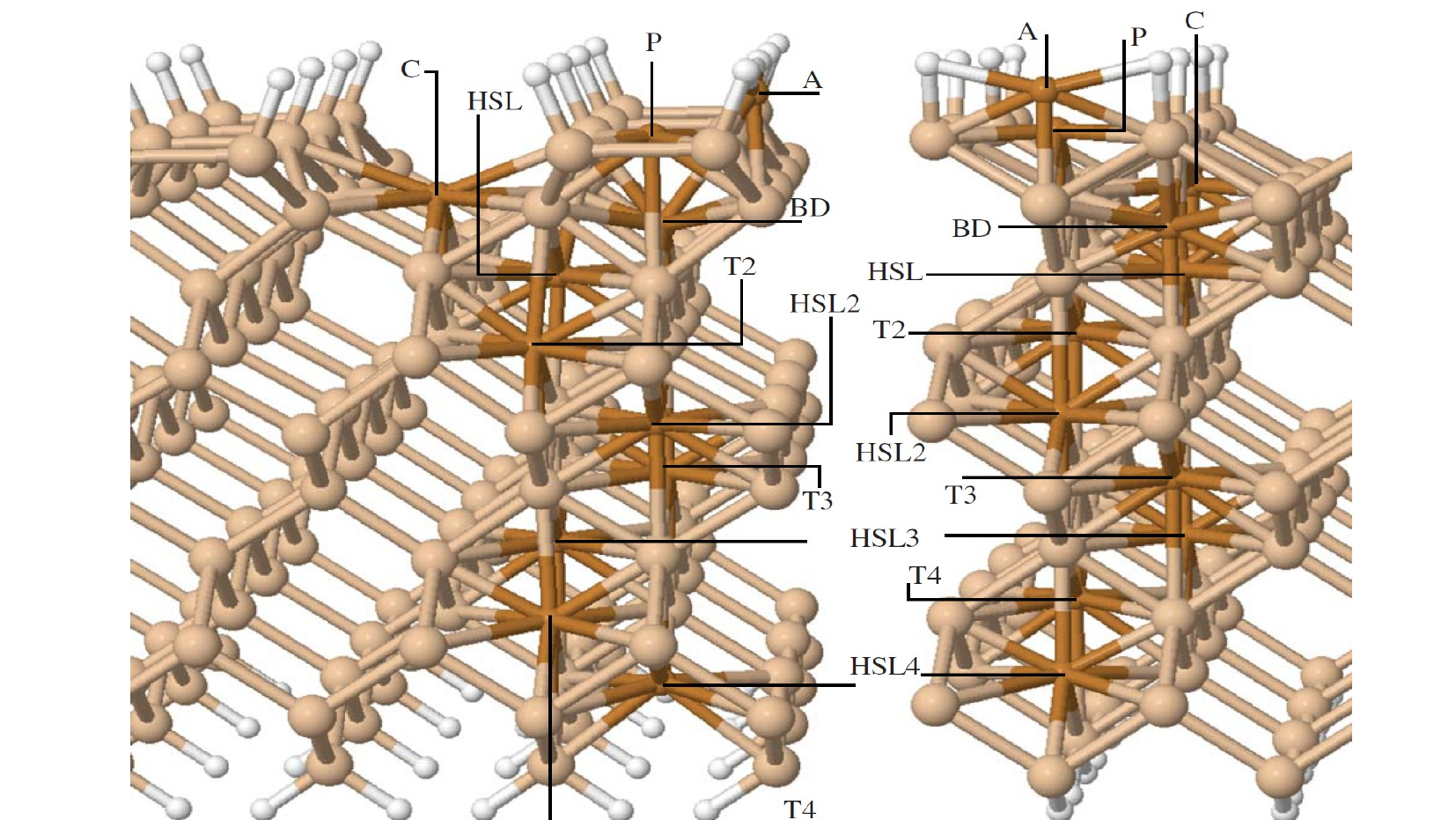}}
\caption{(Color online). Adsorption sites considered within this work both on the surface (P, C, A and B) and beneath (T4, T3, T2, HSL4, HSL3, HSL2, HSL, and BD).}
\label{fig:2}
\end{figure}
%%%%%%%%%%%%%%%%%%%%%%%%%%%%%%%%%%%%%%%%%%%%%%%%%%%%%%%%%%%%%%

\section{Cu diffusion near H-passivated Si(001)}
In this section, we investigate the energy barriers for migration
routes for Cu diffusion. As mentioned above, we will
consider two different Cu sources: Cu deposited from
vacuum and Cu outdiffusing from bulk Si. All the migration
routes considered are shown in Figure 3 (Cu deposited from
vacuum) and Fig. 4 (contaminating Cu inside bulk Si).

\subsection{Depositing Cu from vacuum}
When Cu is deposited from vacuum onto a H-passivated Si(001) surface it will first reach one of the adsorption sites
considered above on the surface: C(-4.41 eV), P(-4.05 eV), and A(-3.96 eV), 
with C being the most feasible site as its BE is the lowest by far. We will show how
adatoms adsorbed at A and P will move to C. These are
illustrated in Fig. 3(a).
A Cu adatom adsorbed at the A site would easily move to
P through a barrier of $E_b$=0.20 eV. Bringing one Cu adatom
between two consecutive P sites along the dimer row direction,
P and P' requires a high energy barrier, Eb=0.71 eV.
The P→BD route crosses a trapezium where the very short
Si-Si bonds (2.43, 2.51, and 3.80 nm) do not give enough
space for the Cu adatom to migrate through; this is the first
example of a migration route labelled “n/a” and described in
Sec. II. In addition, the P→C route, which brings the adtom
to a site whose BE is 0.36 eV better, presents a reasonable
$E_b$=0.44 eV, so that this becomes the most likely route. For
the adatom at C, which is the position most likely to be
reached by deposited Cu adatoms in a first instance, moving
to P is unlikely as C→P route presents a very high
$E_b$=0.80 eV. On the other hand, a possible migration route
C→C' (with no energy gain) along the dimer trench shows a
relatively high $E_b$=0.45 eV, compared to the C→HSL
route, with a very low $E_b$=0.06 eV to a site whose BE is
0.19 eV lower. This indicates essentially free movement of
adatoms from C to HSL. It is interesting to note that previous
studies (Ref. 24) show that Ag adatoms on a H-passivated Si(001)
surface do migrate through the dimer trench, C→C'→C;
this is likely related to the relative sizes of Cu and Ag.
Let us consider now the different migrating posibilities of
Cu adatoms based at HSL (Fig. 3(b)). They could diffuse
towards either BD, C, or T2 sites, since bringing an adatom
from HSL to HSL' directly through the route between the
two Si atoms is not feasible as the Si-Si bond there is so
small. The HSL→BD route has a very high $E_b$=0.83 eV,
whereas bringing the adatom to T2 (a site with the same BE),
HSL→T2, requires only 0.17 eV. Finally, returning to C
(whose BE is 0.19 higher), HSL→C shows an energy barrier
$E_b$=0.19 eV. Therefore, a large number of adatoms will
move to T2 directly. On the other hand while a considerable
fraction of Cu atoms could move to C, they will return to
HSL (with Eb=0.06 eV), and, eventually, diffuse towards
T2. Possible migration routes for adatoms at T2 are displayed
also in Fig. 3(b). A direct route from T2 to T2' would
again find the problem of small Si-Si distances. So the possibilities
for a Cu atom at T2 are either to return to the
hexagonal site HSL', a site with the same BE, through an
energy barrier $E_b$=0.17 eV, or migrating to the hexagonal
site HSL2, with a BE only 0.02 eV lower, through a barrier
almost two times higher, $E_b$=0.28 eV. As BEs of HSL' and
HSL2 differ only 0.02 eV, it is the much lower energy barrier
of the T2→HSL' route which becomes a key factor and will
enhance this route as the most feasible migration route. From
HSL', as we have seen above when dealing with the migration
routes from HSL, Cu adatoms will move towards T2'.
Thus, once adatoms reach T2, they enter a subsurface zigzag
migration route through the sites T2 and HSL, diffusing
along the Si dimer row direction, between the fourth and the
fifth Si layers.
Finally a possible alternative subsurface migration route
along the direction perpendicular to the dimer row, T2→HSL2→T2p (where T2p means the next consecutive
T2 site along the direction perpendicular to the dimer row),
is much less likely due to the higher energy barriers, 0.28
and 0.30 eV, respectively.

%%%%%%%%%%%%%%%%%%%%%%%%%%%%%%%%%%%%%%%%%%%%%%%%%%%%%%%%%%%%%%%%
\begin{figure}[]
\centering
{\includegraphics[width=85mm]{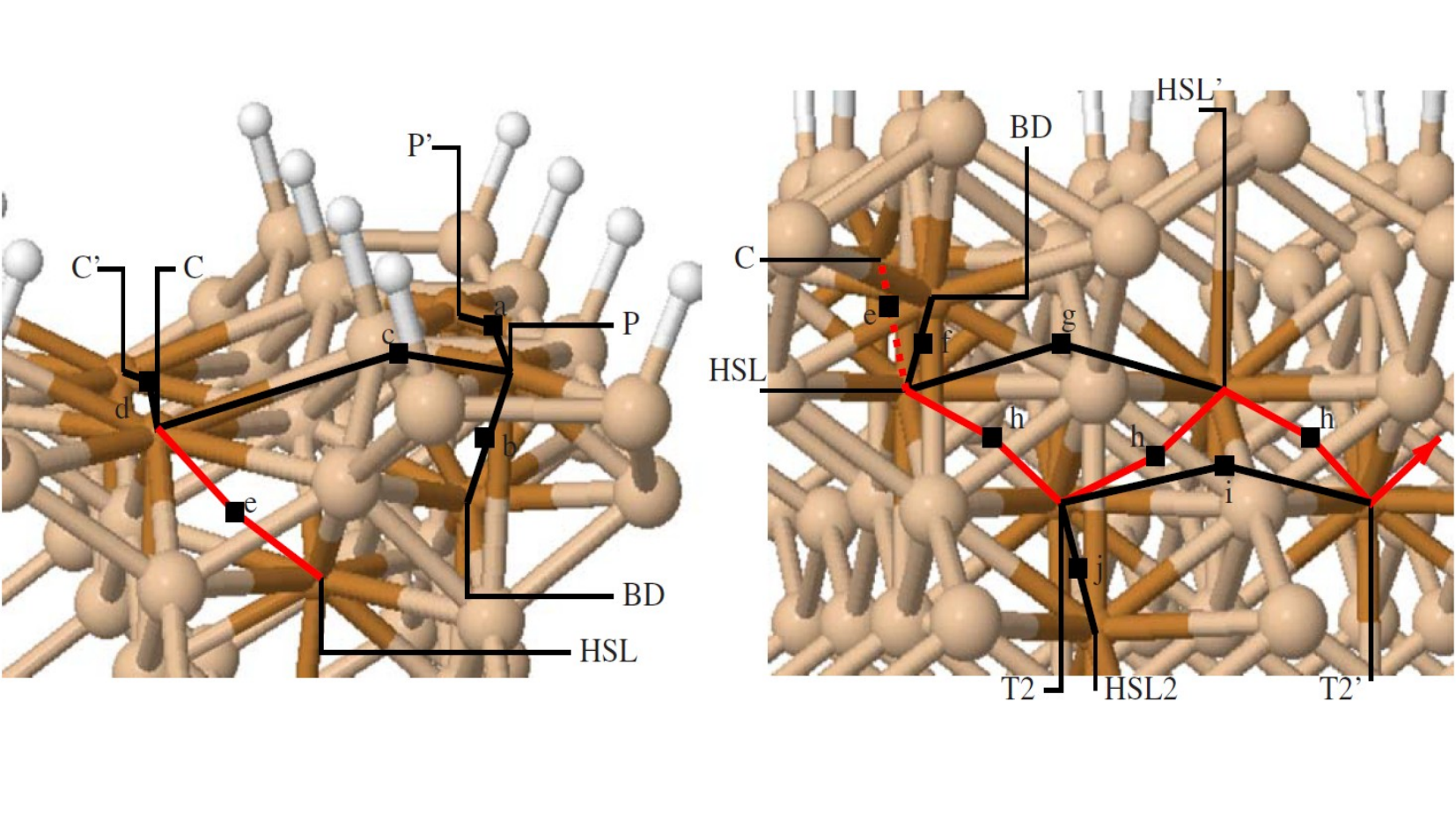}}
\caption{(Color online) Migration routes for Cu deposited from
vacuum: (a) surface routes and (b) subsurface routes. The lowest
energy routes are displayed in red whereas black indicates the rest
of the migration routes considered. Atoms at HSL could also move
towards C where they would return to HSL and would finally migrate
to T2. This fact is reflected in the dashed red line between
HSL and C.}
\label{fig:3}
\end{figure}
%%%%%%%%%%%%%%%%%%%%%%%%%%%%%%%%%%%%%%%%%%%%%%%%%%%%%%%%%%%%%%

\subsection{Contaminating Cu inside bulk Si}
As we have noted above, Cu is known to be a fast diffuser
inside bulk Si and to contaminate Si samples easily. Therefore,
contaminating Cu diffusing through bulk Si might well
be an alternative source of Cu adatoms at the surface. We
have already seen that Cu shows a clear tendency to be adsorbed
close to the surface. In what follows, we investigate
the possible migration routes from bulk to the adsorption
sites near the H-passivated Si(001) surface. We choose HSL4
as our starting point. We note that this is verstudy of subsurface
diffusiony close to the constrained, bulklike layers, but
have confirmed that the diffusion barriers away from this site
are equal to the bulk diffusion barrier (of 0.25 eV), and
conclude that the constraint has only a small effect on
diffusion away from the HSL4 site. It is clear that the
HSL4→HSL4' and HSL4→HSL4p (here HSL4' means the
next consecutive HSL4 site along the dimer row direction,
whereas HSL4p means the next consecutive HSL4 site along
the direction perpendicular to the dimer row) direct route is
not feasible due again to the very small Si-Si bond along this
route. On the other hand, a HSL4→T4 route has a small
barrier $E_b$=0.15 eV, so that Cu moves towards T4, with a
BE 0.12 eV lower than HSL4. Once at T4, Cu could migrate
to either HSL4 or HSL3. The T4→HSL3 is the most feasible
route, as the barrier is relatively small ($E_b$=0.20 eV)
and HSL3 has a BE 0.04 eV lower than T4. However, the
T4→HSL4p route presents a $E_b$=0.27 eV to a site 0.12 eV
higher in BE, which makes possible an alternative zigzag
subsurface migration route T4→HSL4p→T4p along the direction
perpendicular to the dimer row. From HSL3, we consider
possible migrations to T4' and T3. Again, although
more Cu adatoms will follow the HSL3→T3 route, moving
to a site whose BE is 0.12 eV lower through a smaller barrier
$E_b$=0.18 eV, some other adatoms might move towards T4',
with a BE 0.04 eV higher, through a slightly higher barrier
$E_b$=0.25 eV. Thus, despite the tendency of Cu adatoms at
HSL4 and T4 to move towards T3, alternative zigzag migration
routes along both parallel and perpendicular directions
to the dimer row emerge, with the parallel route being more
likely. This alternative migration routes confirm the idea of
fast diffusing Cu.
Once at T3 we check again three possibilities: moving
upwards through HSL2 up to T2, or entering a zigzag subsurface
migration routes either along the dimer row direction,
T3→HSL2→T3', or perpendicular to the dimer row
direction, T3→HSL3p→T3p. The energy barriers are
$E_b$(T3→HSL3)=0.30 eV, $E_b$(T3→HSL2)0.23 eV,
$E_b$(HSL2→T3)=0.20 eV, and $E_b$(HSL2→T2)=0.30 eV,
whereas relative to the BE of the most energetically favoured
adsorption site T3, the BEs of HSL3, HSL2, and T2 are 0.12,
0.03, and 0.05 eV higher. Therefore, Cu adatoms are most
likely to enter the migration route along the dimer row direction,
T3→HSL2→T3'. The perpendicular migration
route is also possible but less likely to happen. Note that the
fraction of Cu adatoms moving along the dimer row direction
has been increased compared to that one at T4 (where
although more adatoms diffused through the dimer row directions
a relative high fraction of Cu could possible migrate
along the perpendicular direction). Finally, the large energy
barrier between HSL2 and T2, $E_b$=0.30 eV, makes the
movement to T2, and entering the T2→HSL→T2' migration
route followed by deposited Cu from vacuum, more unlikely
although kinetically possible. Similarly, although most
of deposited Cu from vacuum follow the T2→HSL→T2'
migration route, a small fraction of adatoms could migrate
from T2 to HSL2 and enter the T3→HSL2→T3' route.
Nevertheless, a vast majority of deposited Cu would migrate
along the dimer row direction. Therefore, this large barrier
prevents a large mixing between deposited Cu from vacuum
and contaminating Cu outdiffusing from bulk Si. We can
conclude that both Cu sources diffuse through different migration
routes, although some mixing can be expected. To
conclude, it is noteworthy that, in spite of multiple diffusion
directions in the botton layers, as we go up, closer to the
surface, Cu diffusion gets more channeled along the dimer
row direction.
In Fig. 5 we summarize the lowest-energy-migration
routes for Cu deposited from vacuum and contaminating Cu
inside bulk Si. Deposited Cu from vacuum follow the route (P→)C→HSL, with $E_b$=0.06 eV, to enter the zigzag
migration route T2→HSL→T2' with $E_b$=0.17 eV.
Contaminating Cu outdiffusing from bulk can diffuse
along different directions, specially when located in the
bottom Si layers. However, we display the most feasible migration
route HSL4→T4→HSL3→T3 to T3, where adatoms
enter a zigzag migration route T3→HSL2→T3'.
Energy barriers are $E_b$(HSL4→T4)=0.15 eV,
$E_b$(T4→HSL3)=0.20 eV, and $E_b$(HSL3→T3)=0.18 eV,
and for the zigzag migration route $E_b$(T3→HSL2)
=0.23 eV and $E_b$(HSL2→T3)=0.20 eV (Table II).

\begin{center}
%\begin{tabular}{ c c c }
\begin{tabular}{  c  c  c  } 
\hline \hline
 Saddle point & Binding energy (eV) & Connecting \\ 
\hline
a & -3.34 & P, P' \\
b & n/a & P, BD \\
c & -3.61 & P, C \\
d & -3.96 & C, C' \\
e & -4.35 & C, HSL \\
f & -3.77 & HSL, BD \\
g & n/a & HSL, HSL' \\
h & -4.43 & HSL, T2 \\
i & n/a & T2, T2' \\
j & -4.32 & T2, HSL2 \\
k & -4.42 & T3, HSL2 \\
l & -4.35 & HSL3, T3 \\
m & -4.29 & T4, HSL3 \\
n & n/a & T4, T4' \\
o & -4.22 & HSL4, T4 \\
p & n/a & HSL4, HSL4'
\end{tabular}
\end{center}

%%%%%%%%%%%%%%%%%%%%%%%%%%%%%%%%%%%%%%%%%%%%%%%%%%%%%%%%%%%%%%%%
\begin{figure}[]
\centering
{\includegraphics[width=95mm]{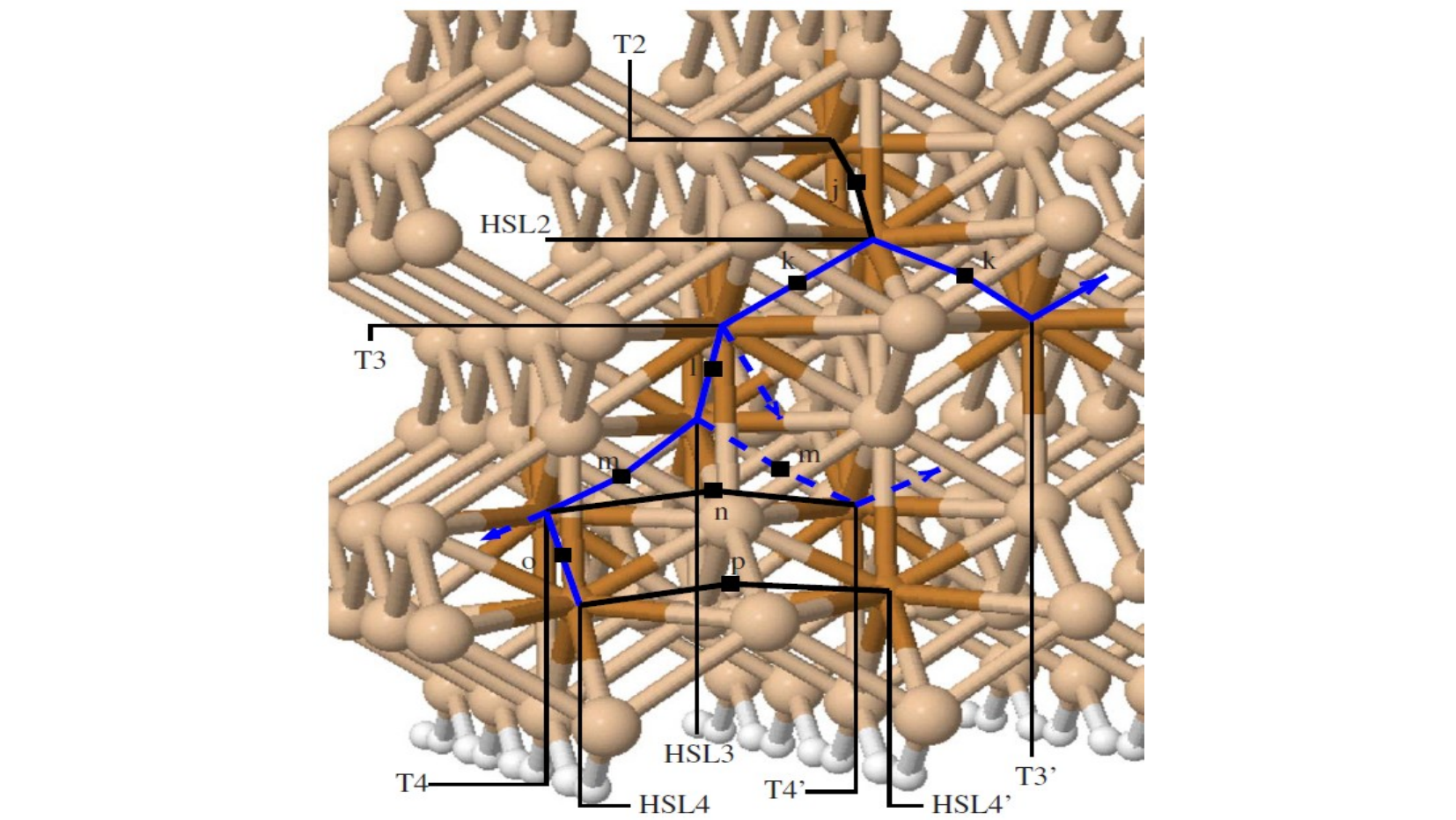}}
\caption{(Color online) Migration routes for contaminating Cu
inside bulk Si. The lowest (possible) energy routes are displayed in
solid (dashed) blue lines whereas black solid lines indicate the rest
of the considered migration routes.}
\label{fig:alpha2}
\end{figure}
%%%%%%%%%%%%%%%%%%%%%%%%%%%%%%%%%%%%%%%%%%%%%%%%%%%%%%%%%%%%%%

%%%%%%%%%%%%%%%%%%%%%%%%%%%%%%%%%%%%%%%%%%%%%%%%%%%%%%%%%%%%%%%%
\begin{figure}[]
\centering
{\includegraphics[width=95mm]{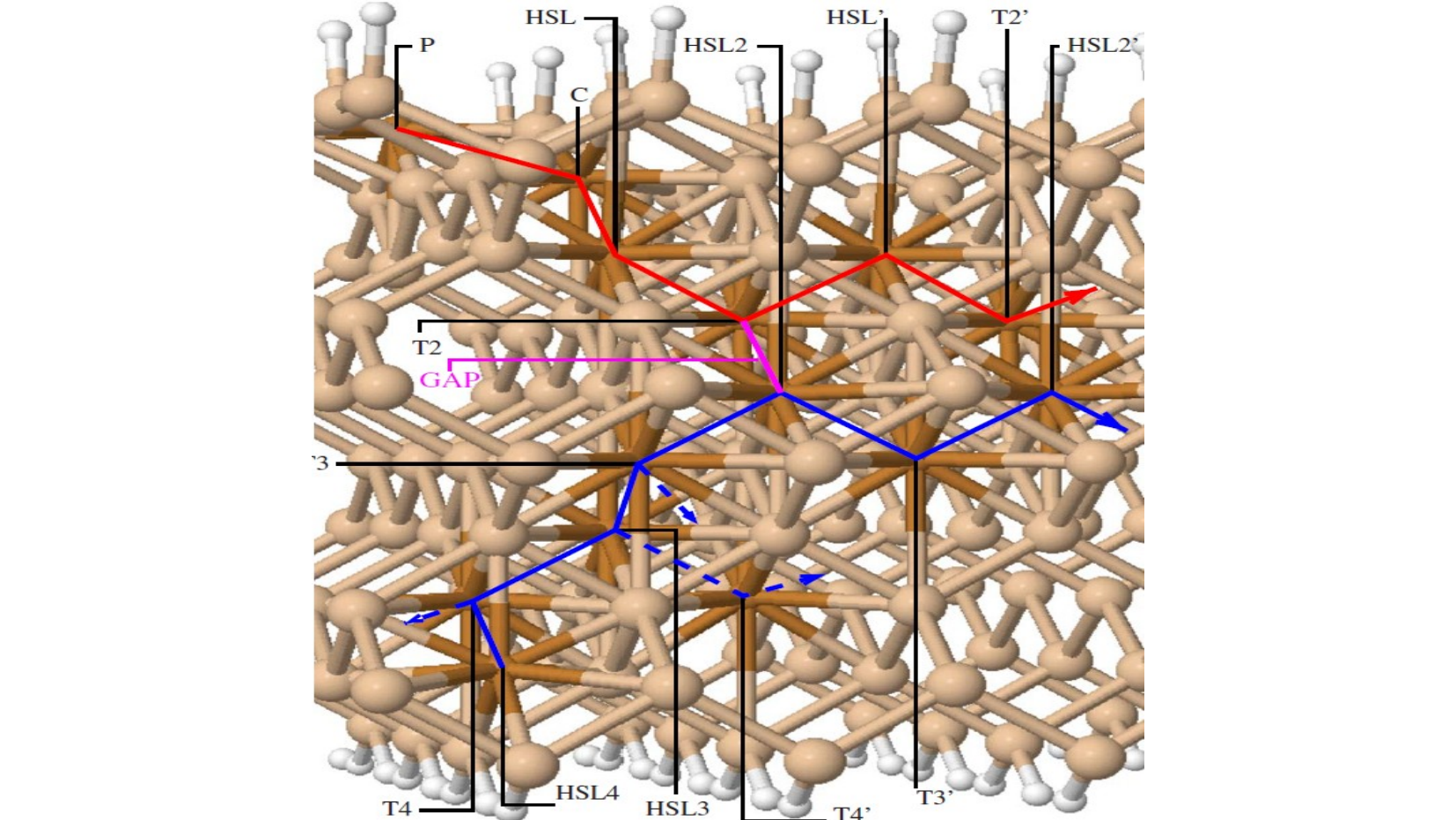}}
\caption{(Color online) Lowest-energy migration routes for both
depositing Cu from vacuum and contaminating Cu from bulk Si.
The migration route for depositing Cu from vacuum is displayed in
red whereas blue is used for contaminating Cu. Alternative possible
migration routes followed by contaminating Cu along both parallel
and perpendicular directions to the dimer row are displayed in blue
dashed lines. Magenta is used to represent the large energy barrier
opened between the T2and HSL2 sites.}
\label{fig:5}
\end{figure}
%%%%%%%%%%%%%%%%%%%%%%%%%%%%%%%%%%%%%%%%%%%%%%%%%%%%%%%%%%%%%%

\section{Results with the one-dimer row slab}
Having established the behaviour of copper in isolation,
we now consider the second unit cell described in Sec. II and
Fig. 1(b), which contains one dimer row with ten Si dimers
and ten layers, and will be suitable for calculations on Bi
nanolines. Our concern is to investigate the influence of the
narrow cell on the Cu energies and barriers.

\subsection{Stable sites for Cu near H:Si(001)}
In Table III we display the BEs of the adsorption sites,
following the same nomenclature as above. As we see,
T3(-5.06 eV) is again the most favourable adsorption site,
followed by HSL2(-5.04 eV), T2(-5.02 eV), and
HSL(-4.99 eV). Although there is a quantitative change in
the BEs, no qualitative difference is observed when compared
with the results obtained with the first slab. The adsorption
positions are ordered exactly in the same way and
the values of $\Delta E (eV)$  are very similar. When using the onedimer
cell one could think that, due to the periodicity along
the direction perpendicular to the dimer row, interactions between
Cu atoms in the neighbouring cell are not neglected.
The fact that there is no qualitative difference at all between
BEs in cells containing one and two dimer rows, respectively,
indicate that this interaction does not play a significant
role. Nevertheless, in what following we will study the energy
barriers and migration routes followed by Cu within the
one-dimer row slab so that we can compare these results to
those obtained earlier. This should give us a better idea on
the issue.
\\
\begin{center}
%\begin{tabular}{ c c c }
\begin{tabular}{  c  c  c  } 
\hline \hline
 Adsorp. site & $\Delta E$ (eV) & Binding energy (eV) \\ 
\hline
 T3 & 0 & -5.06 \\  
 HSL2& 0.02 & -5.04 \\ 
 T2 & 0.04 & -5.02 \\
 HSL& 0.07 & -4.99 \\ 
 HSL3& 0.10 & -4.96 \\
 T4 & 0.20 & -4.86 \\  
 C& 0.21 & -4.85 \\ 
 HSL4 & 0.24 & -4.82 \\
 P& 0.69 & -4.37 \\ 
 A& 0.69 & -4.37 \\
 BD & 0.81 & -4.25 \\
 B & xx & xx
\end{tabular}
\end{center}

\subsection{Depositing Cu from vacuum}
The adsorption sites considered above on the surface
are: C(-4.85 eV), P(-4.37 eV), A(-4.37 eV), and
B(-3.36 eV). As binding energy of B is so bad, we neglect
this site, and consider only the P (A site is very similar) and
C sites. Energy barriers from P(-4.37 eV) to P' and
C(-4.85 eV) are 1.2 and 0.38 eV, respectively, whereas
route towards BD(-4.25 eV) is not feasible due to small
Si-Si bonds. So that Cu adatoms clearly diffuse to C site, as
seen above, though with slightly higher barriers. Energy
barriers from C(-4.85 eV) to C', P(-4.37 eV) and
HSL(-4.99 eV) are 0.53, 0.86, and 0.06 eV, in good agreement
with results from the other cell. Therefore, adatoms
migrate from C to HSL, where possibilities are moving to
BD(-4.25 eV), C, and T2(-5.02 eV), through barriers of
0.45, 0.18, and 0.17 eV. Most of the adatoms will move
directly to T2, while a smaller fraction could go back to C,
from where they will come again to HSL, eventually
diffusing to T2. At T2, possibilities are moving to
HSL'(-4.99 eV) through a 0.20 eV barrier or to
HSL2(-5.04 eV) through a much larger 0.40 eV barrier. It is
clear, then, than Cu adatoms follow a P→C→HSL path to
enter a zigzag subsurface diffusion route along the dimer row direction HSL→T2→HSL'→T2'. This agrees very well
with the migration route calculated with the other slab
model.
\subsection{Outdiffusing Cu from bulk}
In what follows, we investigate the possible migration
routes from bulk to the adsorption sites near the H-passivated
Si(001) surface. We start from HSL4(-4.82 eV), from
where the only possibility is moving to T4(-4.86 eV)
through a 0.19 eV barrier. Energy barrier from T4 to HSL4 is
0.25 eV whereas moving to HSL3(-4.96 eV) requires only
0.19 eV. The latter possibility is, then, more feasible, but a
zigzag route T4→HSL4p→T4p along the direction perpendicular
to the dimer row is also possible; these barriers agree
well with the other cell. From HSL3, most adatoms will
move to T3(-5.06 eV) through a 0.22 eV barrier but a small
fraction of adatoms could move to T4' through a 0.29 eV
barrier. Therefore, despite the tendency of Cu to move from
HSL4 and T4 to T3, alternative zigzag migration routes
emerge along both the parallel and perpendicular directions
to the dimer row, just as it happened when using the former
slab model. At T3 three possibilities emerge: moving upwards
through HSL2(-5.04 eV) to T2(-5.02 eV), or enter
the zigzag migration routes either along the dimer row
direction T3→HSL2→T3', or along the perpendicular
T3→HSL3p→T3p. The energy barriers are
$E_b$(T3→HSL3)=0.32 eV, $E_b$(T3→HSL2)=0.24 eV,
$E_b$(HSL2→T3)=0.22 eV, and $E_b$(HSL2→T2)=0.42 eV,
whereas relative to the BE of the most energetically favoured
adsorption site T3, the BEs of HSL3, HSL2 and T2 are 0.10,
0.02, and 0.04 eV higher. Therefore, Cu adatoms are most
likely to enter the migration route along the dimer row direction,
T3→HSL2→T3'. The perpendicular migration
route is also possible but less likely to happen. Note that the
fraction of Cu adatoms moving along the dimer row direction
has been increased compared to that one at T4 (where
although more adatoms diffused through the dimer row directions
a relative high fraction of Cu could possible migrate
along the perpendicular direction). Again, this facts were
also observed in the previous calculations with the other
slab.
Finally, the large energy barrier between HSL2 and T2,
$E_b$=0.42 eV, makes the movement to T2, and entering the
T2→HSL→T2' migration route followed by deposited Cu
from vacuum, unlikely although kinetically possible. Therefore,
this large barrier prevents a mixing between deposited
Cu from vacuum and contaminating Cu outdiffusing from
bulk Si. This barrier also arises when calculating with a twodimer
rows slab model, although it is slightly smaller.
To summarize: no qualitative difference is observed when
considering slabs containing one and two dimer rows so that
we can conclude considering one-dimer row slab is enough
as the interaction between Cu adatoms situated in neighboring
cells is not significant. Adsorption positions are sorted in
the same way in terms of BE while the most likely migration
routes are identical. Some small changes have been identified,
including a shift in the BE but not in the BE difference
$\Delta E$. Also, when considering the one-dimer row slab, a clear
gap opens between the migration routes followed by Cu deposited
from vacuum and Cu coming from inner layers. Such
an energy gap also emerges in the case of the two dimer rows
slab, although it is smaller and allows a certain mixing between
Cu coming from the two different sources.

\section{Conclusions}
In this paper we have investigated the adsorption and diffusion
of Cu adatoms on the H-passivated Si(001) surface.
We have seen that both the tetrahedral sites T3 and T2, and
the hexagonal sites HSL2 and HSL emerge as the lowest
adsorption energy sites, so that despite the behaviour of Cu
as a fast diffuser inside bulk Si, it tends to be adsorbed close
to the surface. In addition, Cu diffusion for two different
sources of adatoms has been examined. Deposited Cu from
vacuum follow the route (P→)C→HSL, with $E_b$=0.06 eV,
to enter the zigzag migration route T2→HSL→T2' with
$E_b$=0.17 eV. Although contaminating Cu outdiffusing from
bulk can diffuse along different directions, the most feasible
migration route is HSL4→T4→HSL3→T3 to T3, where
adatoms enter a zigzag migration route T3→HSL2→T3'.
Energy barriers are $E_b$(HSL4→T4)=0.15 eV,
$E_b$(T4→HSL3)=0.20 eV, and $E_b$(HSL3→T3)=0.18 eV,
and for the zigzag migration route $E_b$(T3→HSL2)
=0.23 eV and $E_b$(HSL2→T3)=0.20 eV. Therefore, both
Cu starting on the surface (after deposition from vacuum)
and Cu assumed to be contaminating bulk Si diffuse along
the dimer row direction through subsurface zigzag migration
routes. Although contaminating Cu adatoms might diffuse
also along the direction perpendicular to the dimer row when
located in the inner layers, a progressive channeling along
the dimer row direction emerges as adatoms migrate upward
to the surface.

\section{Acknowledgments}
We acknowledge useful discussions with Flemming
Ehlers and James Owen. A. Rodriguez-Prieto. was funded by the Basque
Government. D.R. Bowler was funded by the Royal Society. The
computational work was performed at the London Centre for
Nanotechnology (LCN), UCL.

\label{Bibliography}
\bibliographystyle{unsrt}
\bibliography{Bibliography}

\end{document}